\documentclass[aps,twocolumn,showpacs,preprintnumbers,floatfix]{revtex4}
\usepackage{graphicx}
\usepackage{epsfig}
\usepackage{dcolumn}
\usepackage{amsmath}

\def\be{\begin{equation}}
\def\ee{\end{equation}}
\def\bea{\begin{eqnarray}}
\def\eea{\end{eqnarray}}
\def\bsp{\be\begin{split}}

\def\bes{\be  \begin{split}}

\newcommand{\Rmnum}[1]{\expandafter\@slowromancap\romannumeral #1@}

\begin{document}

\title{Scalar Glueball in Radiative $J/\psi$ Decay on Lattice}
\author{\small
Long-Cheng Gui,$^{1,2}$ Ying Chen,$^{1,2,}$\footnote{cheny@ihep.ac.cn} Gang Li,$^3$ Chuan Liu,$^{4}$ Yu-Bin
Liu,$^{5}$ Jian-Ping Ma,$^{6}$ \\
Yi-Bo Yang,$^{1,2}$ and Jian-Bo~Zhang,$^{7}$\\
(CLQCD Collaboration) } \affiliation{ \small $^1$Institute of High Energy Physics, Chinese Academy of Sciences, Beijing
100049, People's Republic of China \\
$^2$Theoretical Center for Science Facilities, Chinese Academy of Sciences, Beijing 100049,
 People's Republic of China\\
$^3$Department of Physics, Qufu Normal University, Qufu 273165, People's Republic of China\\
$^4$School of Physics and Center for High Energy Physics, Peking University, Beijing 100871, People's Republic of China\\
$^5$School of Physics, Nankai University, Tianjin 300071, People's Republic of China\\
$^6$Institute of Theoretical Physics, Chinese Academy of Sciences, Beijing 100190, People's Republic of China\\
$^7$Department of Physics, Zhejiang University, Zhejiang 310027, People's Republic of China }

\begin{abstract}
The form factors in the radiative decay of $J/\psi$ to a scalar glueball are studied within quenched lattice QCD on
anisotropic lattices. The continuum extrapolation is carried out by using two different lattice spacings. With the
results of these form factors, the partial width of $J/\psi$ radiatively decaying into the pure gauge scalar glueball
is predicted to be $0.35(8)\,{\rm keV}$, which corresponds to a branching ratio of $3.8(9)\times 10^{-3}$. By comparing
with the experiments, our results indicate that $f_0(1710)$ has a larger overlap with the pure gauge glueball than
other related scalar mesons.
\end{abstract}

 \maketitle

The existence of glueballs predicted by QCD remains obscure. For a scalar glueball there is some evidence of its
existence indicated by the fact that there are ten scalar mesons, such as $K^*(1430)$, $a_0(1450)$ and three isoscalars
$f_0(1370)$, $f_0(1500)$, and $f_0(1710)$. These mesons are close in mass and can be sorted into a $SU(3)$ flavor nonet
plus a glueball. Recent lattice studies predict that the lightest pure gauge scalar glueball has a similar
mass~\cite{prd56,prd60,prd73}. Since the scalar glueball can mix with the nearby $q\bar{q}$ mesons, the three isocalars
can be the different admixtures of the pure glueball $G$, the $n\bar{n}$ meson and $s\bar{s}$ meson. So the key problem
is to identify which of the three isoscalars has a dominant glueball component. For this purpose, different mixing
scenarios have been proposed by imposing different mass ordering of $G$, $n\bar{n}$, and $s\bar{s}$ along with the
known decay branching ratios of scalar mesons~\cite{Amsler:1995,
Amsler:1996,Close:2001,Close:2005,Weingarten:1999,Li:2000,Giacosa:2005,Cheng2006}. However, the resultant mixing
patterns are controversial, especially for the status assignment of $f_0(1500)$ and $f_0(1710)$. Obviously, more
theoretical information of the scalar glueball is desired for the problem to be finally resolved.
\par
It is well-known that gluons can be copiously produced in $J/\psi$ decays because of the annihilation of the heavy
quark pair. Among all the decays the radiative decay is of special importance. It is expected that the gluons produced
in $J/\psi$ radiative decays dominantly form a glueball. If the production rate of the scalar glueball in the radiative
decay can be reliably obtained from theoretical studies, it will provide important information for identifying the
possible candidate for the scalar glueball by comparing the production pattern of scalar mesons in these decay
channels. There have been several studies on this topic based on the tree-level perturbative QCD approach and the
dispersion relation method~\cite{Novikov1980, Li1981, Donoghue1983, Liang1993, Meyer2009}, but it is difficult to
estimate theoretical uncertainties in the used approximations. In contrast, lattice QCD provides the rigorous method to
study the radiative decay from first principles. In this Letter, as an exploratory study, we investigate the radiative
decay of $J/\psi$ into a scalar glueball in quenched lattice QCD.

Gauge configurations used in this Letter are generated using the tadpole-improved gauge action~\cite{prd56} on
anisotropic lattices with the temporal lattice spacing much finer than the spatial one, say, $\xi=a_s/a_t=5$, where
$a_s$ and $a_t$ are the spatial and temporal lattice spacing, respectively. Each configuration is separated by 500
heat-bath updating sweeps to avoid the autocorrelation. The much finer lattice in the temporal direction gives a higher
resolution to hadron correlation functions, such that masses of heavy particles can be tackled on relatively coarse
lattices. The calculations are carried out on two anisotropic lattices, namely $L^3\times T=8^3\times 96$ and
$12^3\times 144$. The relevant input parameters are listed in Table~\ref{tab:lattice}, where $a_s$ values are
determined from $r_0^{-1}=410(20)$ MeV. The spatial extension of both lattice is $\sim 1.7\,{\rm fm}$, whose finite
volume effect was found to be small and negligible for glueballs~\cite{prd73}. On the other hand, this lattice size is
large enough for charmonium. For fermions we use the tadpole-improved clover action for anisotropic
lattices~\cite{chuan1}. The parameters in the action are tuned carefully by requiring that the physical dispersion
relations of vector and pseudoscalar mesons are correctly reproduced at each bare quark mass~\cite{chuan2}. The bare
charm quark masses at different $\beta$ are determined by the physical mass of $J/\psi$, $m_{J/\psi}=3.097$ GeV. The
ground state masses of $1S$ and $1P$ charmonia are also calculated with these two lattices (see Fig.~2 and Table II of
Ref.~\cite{Yang:2012} for the details) and the finite $a_s$ effects are found to be small.

\begin{table}[h]
\centering \caption{\label{tab:lattice} The input parameters for the calculation. Values for the
coupling $\beta$, anisotropy $\xi$, the lattice spacing $a_s$, lattice size, and the number of
measurements are listed.}
\begin{ruledtabular}
\begin{tabular}{cccccc}
 $\beta$ &  $\xi$  & $a_s$(fm) & $La_s$(fm)&
 $L^3\times T$ & $N_{\rm conf}$ \\\hline
   2.4  & 5 & 0.222(2) & 1.78 &$8^3\times 96$ & 5000 \\
   2.8  & 5 & 0.138(1) & 1.66 &$12^3\times 144$ & 5000  \\
\end{tabular}
\end{ruledtabular}
\end{table}
\par
\begin{figure}
\includegraphics[height=4.5cm]{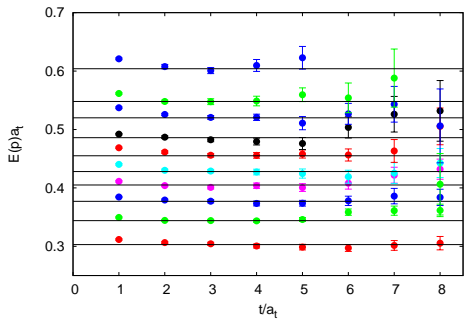}
\caption{The effective energy plot for the $A_1^{++}$ glueball with different spatial momenta. From
top to bottom are the plateaus for momentum modes, $\vec{p}=2\pi\vec{n}/L$, with $\vec{n}=(2,2,2)$,
$(2,2,1)$, $(2,2,0)$,
 $(2,1,1)$, $(2,1,0)$, $(2,0,0)$, $(1,1,1)$, $(1,1,0)$, $(1,0,0)$, and $(0,0,0)$. \label{glb_plat}}
\end{figure}
\par
To lowest order in QED, the amplitude $M$ for radiative decay
$J/\psi\rightarrow\gamma G$ is given by
\begin{equation}
M_{r,r_\gamma}=\epsilon_{\mu}^*(\vec{q},r_\gamma)\langle
G(\vec{p}_f)|j^{\mu}(0)|J/\psi(\vec{p}_i,r)\rangle,
\end{equation}
where $\vec{q}=\vec{p}_i-\vec{p}_f$ is the momentum of the real photon, $r$ and $r_\gamma$ are the quantum numbers of
the polarizations of $J/\psi$ and the photon, respectively. $\epsilon(\vec{q},r_\gamma)$ is the polarization vector of
the photon and $j^\mu$ is the electromagnetic current operator. The hadronic matrix element appearing in the above
equation can be obtained directly from lattice QCD calculation of corresponding three-point functions.
\par
One of the key issues in our calculation is to construct the
interpolating field operator which couples dominantly to the
so-called pure gauge scalar glueball, which is defined by using
interpolating field operators built from the gauge fields only. For
this purpose, we adopt the variational method along with the
single-link and double-link smearing schemes~\cite{prd60,prd73}.
More specifically, since the irreducible representation $A_1^{++}$
of lattice symmetry group $O$ gives the right quantum number
$J^{PC}=0^{++}$ in the continuum limit, we construct an $A_1^{++}$
operator set $\{\phi_\alpha, \alpha = 1,2,\ldots, 24\}$ of 24
different gluonic operators. Through the Fourier transformation,
\begin{equation}
\phi_\alpha(\vec{p},t)=\sum\limits_{\vec{x}}\phi_\alpha(\vec{x},t)e^{-i\vec{p}\cdot \vec{x}},
\end{equation}
we obtain the operator set $\{\phi_\alpha(\vec{p},t), \alpha = 1,2,\ldots, 24\}$ which couples to an $A_1^{++}$
glueball state with the definite momentum $\vec{p}$. For each $\vec{p}$, by solving the generalized eigenvalue problem,
\begin{equation}
\label{eigen} \tilde{C}(t_D){\bf v}^{(R)} = e^{-t_D\tilde{m}(t_D)}\tilde{C}(0){\bf v}^{(R)},
\end{equation}
at $t_D=1$, where $\tilde{C}(t)$ is the correlation matrix of the operator set,
\begin{equation}
\tilde{C}_{\alpha\beta}(t) =\frac{1}{N_t} \sum\limits_{\tau}\langle
0|{\phi}_\alpha(\vec{p},t+\tau){\phi}_\beta^\dagger (\vec{p},\tau)|0\rangle,
\end{equation}
we obtain an optimal combination of the set of operators, $\Phi(\vec{p},t)=\sum
v_{\alpha}\phi_\alpha(\vec{p},t)$, which overlaps most to the ground state,
\begin{eqnarray}
\label{glb_two}
 C(\vec{p},t)&=&\frac{1}{T}\sum\limits_{\tau}\langle
\Phi(\vec{p},t+\tau)\Phi^\dagger(\vec{p},\tau)\rangle \nonumber\\
&\approx& \frac{|\langle 0|\Phi(\vec{p},0)|S(\vec{p})\rangle|^2}{2E_SV_3}e^{-E_St}\approx e^{-E_St},
\end{eqnarray}
where the normalization $C(\vec{p},0)=1$ is also used. This is actually the case that $C(t)$ can be well described by a
single exponential, $C(t)=We^{-Et}$, with $W$ usually deviating from one by few percents. Figure~\ref{glb_plat} shows
the effective energy plateaus of the $A_1^{++}$ glueball for typical momentum modes, where one can see that the
plateaus start even from $t=1$.

Glueballs are noisy objects and large statistics is usually required. In this Letter, we generated 5000 configurations
for both lattice systems. In order to increase the statistics additionally, for each configuration we calculate $T$
charm quark propagators $S_F(\vec{x},t;\vec{0},\tau)$ by setting a point source on each time slice $\tau$, which
permits us to average over the temporal direction when calculating the three-point functions. Therefore, the
three-point functions we calculate in this Letter are
\begin{widetext}
\begin{eqnarray}
\Gamma_{\mu,j}^{(3)}(\vec{p}_f,\vec{q};t_f,t) &=& \frac{1}{T}\sum\limits_{\tau=0}^{T-1}\sum\limits_{\vec{y}}
e^{-i\vec{q}\cdot \vec{y}} \langle
\Phi(\vec{p}_f,t_f+\tau)J_\mu (\vec{y},t+\tau)O_{V,j}(\vec{0},\tau)\rangle\nonumber\\
&=& \frac{1}{T}\sum\limits_{\tau=0}^{T-1}\sum\limits_{\vec{y}} e^{-i\vec{q}\cdot \vec{y}} \left\langle
\Phi(\vec{p}_f,t_f+\tau){\rm Tr}\left[\gamma_\mu S_F(\vec{y},t+\tau;\vec{0},\tau)\gamma_j\gamma_5
S_F^\dagger(\vec{y},t+\tau;\vec{0},\tau)\gamma_5\right]\right\rangle\nonumber\\
&=&\sum\limits_{S,V,r}\frac{e^{-E_S (t_f-t)}e^{-E_V
t}}{2E_S(\vec{p}_f)V_3 2E_V(\vec{p}_i)} \langle
0|\Phi(\vec{p}_f,0)|S(\vec{p}_f)\rangle \langle
S(\vec{p}_f)|J_\mu(0)|V(\vec{p}_i,r)\rangle\langle
V(\vec{p}_i,r)|O_{V,j}^{\dagger}(0)|0\rangle,\nonumber\\
\end{eqnarray}
\end{widetext}
where $J_\mu(x) =\bar{c}(x)\gamma_\mu c(x)$ is the vector current operator, $O_{V,j}=\bar{c}\gamma_j c$ the
conventional interpolation field for $J/\psi$, and the summation in the last equality is over all the possible states
and vector polarizations, $\vec{p}_i$ is the spatial momentum of the initial vector charmonium and satisfies the
relation $\vec{p}_i=\vec{p}_f+\vec{q}$. The vector current $J_\mu (x)$, which is conserved in the continuum limit, is
no longer conserved on the lattice which requires a multiplicative renormalization. In this Letter, we adopt the
nonperturbative strategy proposed by Ref.~\cite{dudek06} to define the renormalization constant,
\begin{equation}
Z_V^\mu(t)=\frac{p^\mu}{2E(p)}\frac{\Gamma^{(2)}_{\eta_c\eta_c}(\vec{p},t_f=T/2)}
{\Gamma^{(3)}_{\eta_c\gamma_\mu \eta_c}(\vec{p}_f=\vec{p}_i=\vec{p}, t_f=T/2,t)},
\end{equation}
where $\Gamma^{(2)}_{\eta_c\eta_c}$ is the two-point function of the pseudoscalar charmonium $\eta_c$,
$\Gamma^{(3)}_{\eta_c\gamma_\mu \eta_c}$ is the corresponding three point function with the vector current insertion on
one of the quark lines. It should be remarked that the possible disconnected diagrams due to the charm and
quark-antiquark annihilation are neglected in this Letter.
\par
The parameters $E_S$, $E_V$, the matrix elements $\langle 0|\Phi(\vec{p}_f,0)|S(\vec{p}_f)\rangle$ and $\langle
0|O_{V,j}|V(\vec{p}_i,r)\rangle$ can be derived from the relevant two-point functions of glueballs and $J/\psi$.
Specifically, from Eq.~(\ref{glb_two}) we have
\begin{equation}
\langle 0|\Phi(\vec{p}_f,0)|S(\vec{p}_f)\rangle\approx
\sqrt{2E_S(\vec{p}_f)V_3}.
\end{equation}
For the vector meson we take the following convention,
\begin{equation}
\label{psi_decay_constant}
 \langle 0|O_{V,j}(0)|V(\vec{p},r)\rangle = f_V \epsilon_j(\vec{p},r),
\end{equation}
where $f_V$ is a parameter independent of $\vec{p}$, and $\epsilon_j(\vec{p},r)$ the polarization vector of the vector
meson, whose concrete expression depends on reference frames and is irrelevant to the calculation in this Letter. By
using the multipole decomposition, the matrix elements $\langle S(\vec{p}_f)|J_\mu(0)|V(\vec{p}_i,r)\rangle$ can be
written as~\cite{dudek06},
\begin{equation}
\sum_r \langle S(\vec{p}_f)|J_\mu(0)|V(\vec{p}_i,r)\rangle
\epsilon_j(\vec{p}_i,r)=\alpha_{\mu j}E_1(Q^2)+\beta_{\mu
j}C_1(Q^2),
\end{equation}
where $\alpha_{\mu j}$ and $\beta_{\mu j}$
 are known functions of $p_f$ and $p_i$ (their explicit
expressions are neglected here), $E_1(Q^2)$ and $C_1(Q^2)$ are the two form factors which depend only on
$Q^2=-(p_i-p_f)^2$. Only the form factor $E_1(Q^2)$ will be needed to determine the decay width with
\begin{equation}\label{scalar_width}
\Gamma(J/\psi\rightarrow \gamma
G_{0^{++}})=\frac{4}{27}\alpha\frac{|\vec{p}_\gamma|}{M_{J/\psi}^2}|E_1(0)|^2,
\end{equation}
where $\alpha$ is the fine structure constant, $p_\gamma$ the photon
momentum with $|\vec{p}_\gamma|=(M_{J/\psi}^2-M_G^2)/(2M_{J/\psi})$.
Therefore we will only focus on the extraction of $E_1(Q^2)$.
\par
To measure $E_1(Q^2)$ with different $Q^2$ on the lattice, we create
$J/\psi$ on lattices with the momentum $\vec{p}_i=\vec{0}$ or
$|\vec{p}_i|=2\pi/La_s$,
 and the scalar glueball with the momentum $\vec{p}_f=2\pi \vec{n}/La_s$, where  $\vec{n}$ is ranged
from $(0,0,0)$ to $(2,2,2)$. Among all the combinations of the vector current index $\mu$, the polarization index $j$,
the glueball momentum $p_f$ and the $J/\psi$ momentum $p_i$, it is found that there are specific combinations which
give $\alpha_{\mu i}(p_f,p_i)=1$ and $\beta_{\mu i}(p_f,p_i)=0$. Hereafter, we will only select these combinations for
our practical data analysis. An additional benefit of this selection is that in these combinations one has $\sum_r
\epsilon_j^*(\vec{p}_i,r)\epsilon_j(\vec{p}_i,r)=1$.

\par
With these prescriptions, the form factor $E_1(Q^2)$ can be derived as,
\begin{eqnarray}
\tilde{E}_1(Q^2,t_f,t)&\approx& \frac {Z_V^{(s)} \Gamma^{(3)}(\vec{p_f},\vec{p}_i;t_f,t)}
{C(\vec{p}_f,t_f-t)\Gamma^{(2)}(\vec{p}_i,t)}f_V \sqrt{2E_S(\vec{p}_f)V_3}\nonumber\\
&&(t,t_f-t\gg 0),
\end{eqnarray}
where $Q^2$ can be given by $\vec{p}_i$ and $\vec{p}_f$, the indices of the three-point function $\Gamma^{(3)}$ and the
related two-point functions $\Gamma^{(2)}$ are omitted here, and $Z_V^{(s)}$ is the renormalization constant of the
spatial components of the vector current. In practice, the symmetric indices and momentum combinations which give the
same $Q^2$ are averaged to increase the statistics. Traditionally, the time separation $t$ and $t_f-t$ should be kept
large enough for the ground states to contribute dominantly to the three point function. Even with this large
statistics, we find that the signal of the glueball damps rapidly with respect to $t_f-t$. However, this is not a real
disaster since the optimal glueball operators we use couple almost exclusively to the ground state, as is mentioned
before. So we fix $t_f-t=1$ with varying $t$ and extract $E_1(Q^2)$ from the plateaus of $\tilde{E}_1(Q^2,t_f,t)$. With
the very high statistics in this Letter, the hadron parameters, such as the energies of the glueball and $J/\psi$, the
constant $f_V$ in Eq.~(\ref{psi_decay_constant}) and the matrix elements $\langle 0|\Phi(\vec{p},0)|S(\vec{p})\rangle$
can be determined very precisely and are treated as known parameters.
\begin{figure}[t!]
\includegraphics[height=4cm]{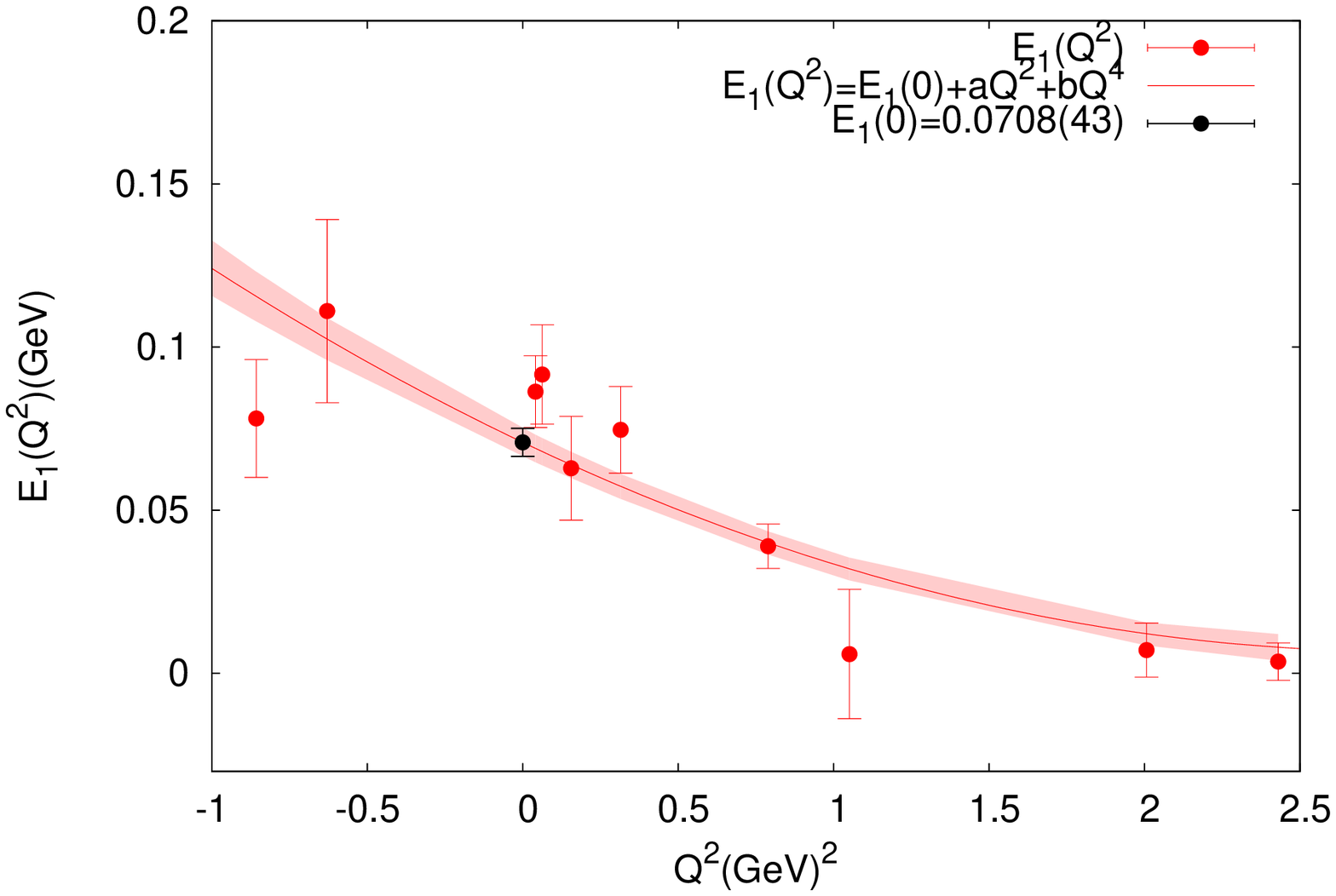}
\includegraphics[height=4cm]{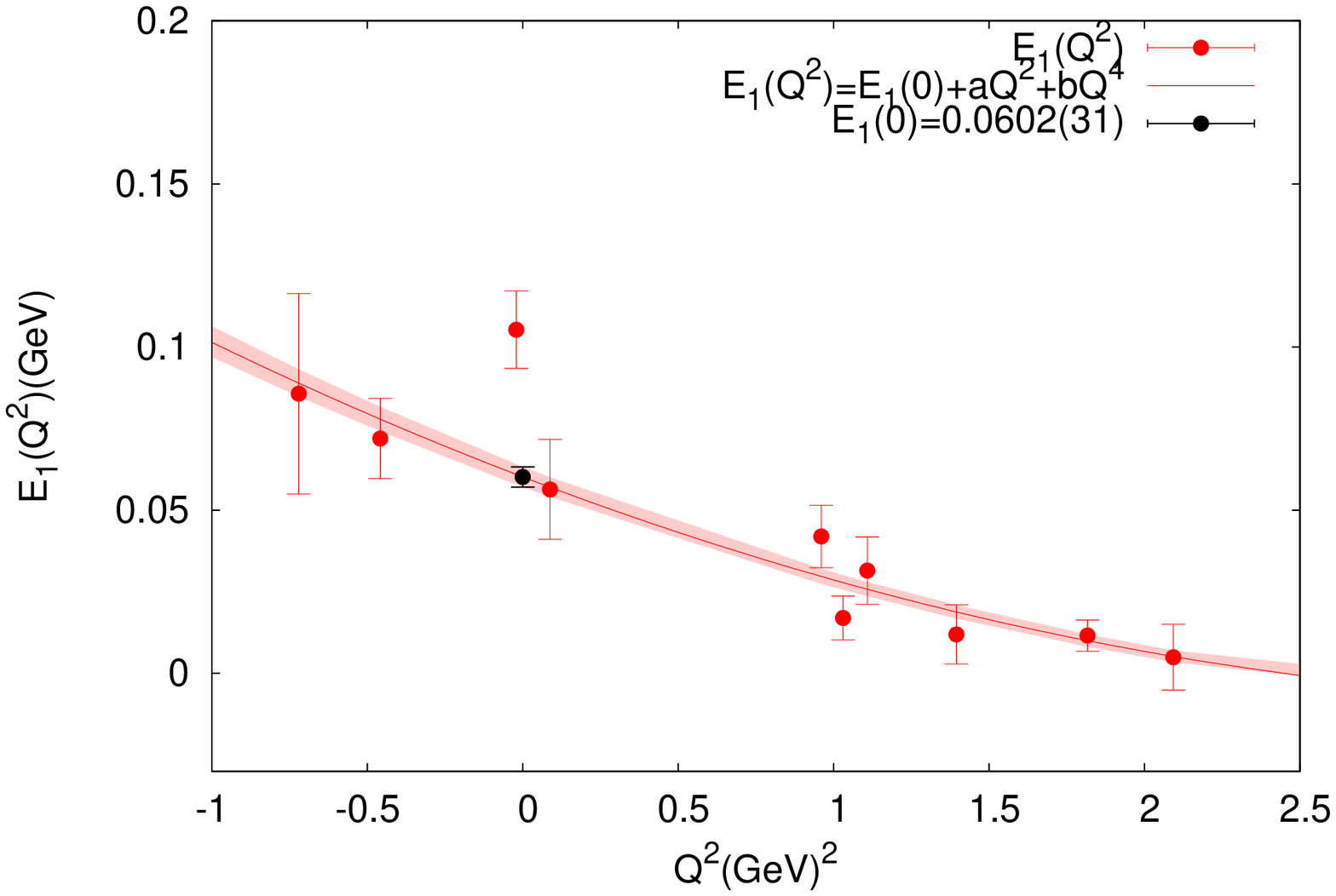}
\caption{\label{form}The extracted form factors $E_1(Q^2)$ in the physical units. The upper panel is for $\beta=2.4$
and the lower one for $\beta=2.8$. The curves with error bands show the polynomial fit with
$E_1(Q^2)=E_1(0)+aQ^2+bQ^4$, as the black dot is the interpolated value $E_1(0)$ at $Q^2=0$.}
\end{figure}
\par
$E_1(Q^2)$ for different $Q^2$ are extracted from the same configuration ensemble and are therefore highly correlated.
In the data analysis we fit them through the correlated data fitting. For each lattice system, the 5000 configurations
are divided into 100 bins with 50 configurations in each bin. The measurements in each bin are averaged and the average
is taken as an independent measurement. After that, all  $E_1(Q^2)$s are extracted simultaneously through the jackknife
method. In order to get the form factor at $Q^2=0$, we carry out a correlated polynomial fit to the $E_1(Q^2)$ from
$Q^2=-1.0 \,{\rm GeV}^2$ to $2.5 \,{\rm GeV}^2$,
\begin{equation}
E_1(Q^2)=E_1(0)+aQ^2+bQ^4.
\end{equation}
Figure~\ref{form} shows the final results of $E_1(Q^2)$ for $\beta=2.4$ (left panel) and $\beta=2.8$ (right panel),
where the red points are the calculated value with jackknife errors, and the red curves are the polynomial fit with
jackknife error bands, the black points label the interpolated $E_1(0,a_s)$.
\par
The last step is the continuum extrapolation using the two lattice systems. The continuum limit of $E_1(0,a_s)$ is
determined to be $E_1(0)=0.0536(57)\,{\rm GeV}$ by performing a linear extrapolation in $a_s^2$. For the continuum
value of the scalar glueball mass, we take $M_G=1.710(90)\,{\rm GeV}$ from Ref.~\cite{prd73}. Thus, according to
Eq.~(\ref{scalar_width}), we finally get the decay width $\Gamma(J/\psi\rightarrow \gamma G_{0^{++}})=0.35(8)\,{\rm
keV}$. Using the reported total width of $J/\psi$, $\Gamma_{\rm tot}=92.9(2.8)\,{\rm keV}$, the corresponding branching
ratio is
\begin{equation}
\Gamma(J/\psi\rightarrow \gamma G_{0^{++}})/\Gamma_{\rm tot}=3.8(9)\times 10^{-3}.
\end{equation}

\begin{table}[t]
\centering \caption{\label{twobeta} Listed in the table are the $A_1^{++}$ glueball masses $M_G$, the renormalization
constants $Z_V^{(s)}(a_s)$ of the spatial component of the vector current, and the form factors $E_1(Q^2=0,a_s)$
calculated on the two lattices with $\beta=2.4$ and 2.8, respectively. Also shown are the continuum extrapolation of
$E_1(0)$ and the resultant partial width $\Gamma$.}
\begin{ruledtabular}
\begin{tabular}{ccccc}
 $\beta$  &  $M_G$(GeV) & $Z_V^{(s)}(a_s)$  &  $E_1(0,a_s)$ (GeV) & $\Gamma$(keV)  \\
 \hline
   2.4   &  1.360(9)   & 1.39(2)   &    0.0708(43)     &    \ldots    \\
   2.8   &  1.537(7)   & 1.11(1)   &    0.0602(31)     &    \ldots    \\
   $\infty$ & 1.710(90)~\cite{prd73}& \ldots       &    0.0536(57)     &    0.35(8) \\
\end{tabular}
\end{ruledtabular}
\end{table}
\par
By comparing our result with their production rates in the radiative decay of $J/\psi$, we can get some useful
information for the glueball components of the scalar mesons $f_0(1710)$, $f_0(1500)$, and $f_0(1370)$. From
PDG2010~\cite{PDG2010}, the branching ratios of the observed radiative decay modes of $J/\psi$ to $f_0(1710)$ are:
$Br(J/\psi\rightarrow \gamma f_0(1710)\rightarrow \gamma K\bar{K})=8.5_{-0.9}^{+1.2}\times 10^{-4}$,
$Br(J/\psi\rightarrow \gamma f_0(1710)\rightarrow \gamma \pi\pi)=(4.0\pm 1.0)\times 10^{-4}$, $Br(J/\psi\rightarrow
\gamma f_0(1710)\rightarrow \gamma \omega\omega)=(3.1\pm 1.0)\times 10^{-4}$, which add up to about $1.5\times
10^{-3}$. With  the measured branching ratio $Br(f_0(1710)\rightarrow K\bar{K})=0.36\pm 0.12$~\cite{Albaladejo08}, and
the ratio $\Gamma(f_0(1710)\rightarrow\pi\pi)/\Gamma(f_0(1710)\rightarrow
K\bar{K})=0.41_{-0.17}^{+0.11}$~\cite{Ablikim06} (Ref.~\cite{Albaladejo08} also predicts this ratio to be $0.32\pm
0.14$ from a coupled channel study of meson-meson $S$-waves), one can estimate the production rate of $f_0(1710)$ to be
$(2.4\pm 0.8)\times 10^{-3}$ or $(2.7\pm 1.3)\times 10^{-3}$. This is compatible with our lattice result. For the
$f_0(1500)$, PDG2010 gives a lower bound to its production rate in $J/\psi$ radiative decay, $Br(J/\psi\rightarrow
\gamma f_0(1500))
>5.7(8)\times 10^{-4}$~\cite{PDG2010}. On the other hand, with the BESII result
$Br(J/\psi\rightarrow \gamma f_0(1500)\rightarrow \gamma \pi\pi)=(1.01\pm 0.32)\times 10^{-4}$~\cite{Ablikim06}, and
the branching ratio $Br(f_0(1500)\rightarrow\pi\pi)=0.349\pm{0.023}$~\cite{PDG2010}, $Br(J/\psi\rightarrow \gamma
f_0(1500))$ is estimated to be $2.9(9)\times 10^{-4}$. Both are much smaller than our prediction. Finally, there is no
evidence of the production of $f_0(1370)$ in the $J/\psi$ radiative decays. Based on this comparison, $f_0(1710)$ seems
to have scalar glueballs as its dominant components, while for the other two scalar mesons, this does not seem to be
the case.
\par
To summarize, we have carried out the first lattice study on the $E_1$ amplitude of $J/\psi$ radiatively decays into
the pure gauge scalar glueball $G_{0^{++}}$ in the quenched approximation. With two different lattice spacings, the
decay amplitude is extrapolated to the continuum limit with a value $E_1(Q^2=0)=0.0536(57)\,{\rm GeV}$. Thus, the
partial decay width $\Gamma(J/\psi\rightarrow \gamma G_{0^{++}})$ is predicted to be $0.35(8)\,{\rm keV}$, which gives
the branching ratio $\Gamma/\Gamma_{\rm tot}=3.8(9)\times 10^{-3}$. We admit that the systematic uncertainties due to
the quenched approximation are not under control in this Letter, however, we are pleased to see that a recent
$2+1$-flavor dynamical lattice study on the glueball spectrum claims that there are not large unquenching effects
observed, especially for the scalar and tensor glueballs~\cite{Gregory:2012}. Anyway, our results are helpful in the
sense that $f_0(1710)$ appears to have the largest overlap to the pure gauge glueball among the relevant scalar mesons.
We hope this fact sheds some light on the long-lasting puzzle of the identification of the scalar glueball.

We acknowledge fruitful discussions with Keh-Fei Liu and Hai-Yang Cheng. This work is supported in part by the National
Science Foundation of China (NSFC) under Grants No.10835002, No.11075167, No.11021092, No. 10675101, No. 10947007, and
No.10975076. Y. C and C. L also acknowledge the support of NSFC and DFG (CRC110).

\end{document}